\documentclass[a4,11pt]{iopart}

\usepackage{amstext}
\usepackage{amssymb}
\usepackage{iopams}

\usepackage{epsfig}
\usepackage{graphics}
\usepackage{graphicx}
\newcommand{\dd}{{\rm d}}

\usepackage{graphicx,xcolor}
\usepackage[latin1]{inputenc}
\usepackage[T1]{fontenc}
\usepackage[french,german,english]{babel}
\usepackage{ae,aecompl}

\newcommand{\smallmatrix}[1]{
  \text{\begin{footnotesize}
      $ \pmatrix{ %
        #1} $
    \end{footnotesize}}}

\newcommand{\eqref}[1]{(\ref{#1})}

\usepackage[pdftex]{hyperref}

\begin{document}

\title[Phase diagram of two-lane driven diffusive systems]
{Phase diagram of two-lane driven diffusive systems}

\author{M.~R.~Evans$^1$, Y.~Kafri$^2$, K.~E.~P.~Sugden$^1$, J.~Tailleur$^1$}

\address{$^1$ SUPA, School of Physics and Astronomy, University of
  Edinburgh, Mayfield Road, Edinburgh EH9 3JZ, Scotland}

\address{$^2$ Department of Physics, Technion, Haifa, 32000, Israel}

\eads{\mailto{martin@ph.ed.ac.uk}, 
\mailto{kafri@physics.technion.ac.il},
\mailto{Julien.Tailleur@ed.ac.uk},
\mailto{ksugden@ed-alumni.net}}
  \date{today}
  \begin{abstract}
We consider a large class of two-lane driven diffusive systems
{in contact with reservoirs at their boundaries} and develop a
  stability analysis as a method to derive the phase diagrams of such
  systems.  We illustrate the method by deriving phase diagrams for
  the asymmetric exclusion process coupled to various second lanes: a
  diffusive lane; an asymmetric exclusion process with advection in
  the same direction as the first lane, and an asymmetric exclusion
  process with advection in the opposite direction.  The competing
  currents on the two lanes naturally lead to a very rich
  phenomenology and we find a variety of phase diagrams.  It is shown
  that the stability analysis is equivalent to an `extremal current
  principle' for the total current in the two lanes.  We also point to
  classes of models where both the stability analysis and the extremal
  current principle fail.
\end{abstract}
  
  \maketitle

\section{Introduction}

The physics of many non-equilibrium processes ranging from ionic
conductors~\cite{KLS} and spin transport~\cite{RFF06} to biological
transport~\cite{MGP68,AMP99,GHS08,SEPR07,ES07,CKZCJ06} can be
described by driven diffusive systems.  These are systems held out of
equilibrium, for instance, by the environment driving a current of
particles or energy through them.  In the long-time limit, these
systems can attain stationary states that are distinct from equilibrium
states, and that escape description by conventional statistical mechanics.
These systems have been widely studied for their fascinating behaviour, in
particular, their propensity to exhibit boundary-induced phase
transitions \cite{Mukamel00,SZ,Krug91}.

Much work on driven diffusive systems has been carried out in one
dimension---one-dimensional systems being of particular interest as
they exhibit phase transitions with no equilibrium counterparts. A
paradigmatic example is the totally asymmetric simple exclusion
process (TASEP) whose phase diagram for the open boundary case has
been studied in detail. Although in special cases such as the open
boundary TASEP exact solutions are available \cite{DEHP93,SD93,BE07},
a useful approach for more general driven systems is to construct a
mean-field theory. Remarkably, for the open boundary TASEP a simple
mean-field approximation recovers the exact phase
diagram~\cite{Mukamel00,DDM02,MB05}. Building on the mean-field
theory, an ``extremal current principle'' has been proposed.  This
principle allows one to construct phase diagrams of one-dimensional
open driven systems without explicitly solving the, generally
non-linear, mean-field equations~\cite{Krug91,PS99,HKPS01}.

More recently there has been much interest in extensions of the
one-dimensional case to multiple lanes.  A need for additional lanes
arises naturally when describing many classes of systems.  As
an example, consider a biological transport system which describes the
motion of molecular motors along a protofilament. Since molecular
motors have a finite processivity they eventually detach and diffuse
in the environment. To model the environment in a simplistic manner, a
second lane in which the particles are diffusive and experience no
exclusion can be introduced~\cite{KL03}. (For a different approach to
tackle the same problem see~\cite{PFF03,NOSC05,GGNSC}.) Two-lane
models have also been used to describe the extraction of membrane
tubes by molecular motors~\cite{TEK09}, macroscopic clustering
phenomena~\cite{KSZ99}, spin transport~\cite{MRFF10} and various
systems of oppositely moving
particles~\cite{SKZ05,Juhasz07,JNHWW09,Juhasz10}.  In particular
systems consisting of two coupled TASEPs have been the subject of several
studies~\cite{PK04,PK06,HS05,RFF07,JHWW08,SARS10}.  More general
multilane systems have also been studied~\cite{PS04,CGW08} and the
hydrodynamics of coupled two-species systems has been considered
in~\cite{PS03,Schutz03,Mukherji09}.

Given the broad applicability and interest in multilane driven
diffusive models it is interesting to ask if there are generic
principles, akin to the extremal current principle, which can be
used to construct phase diagrams of such systems. In this paper we
address this question for open boundary two-lane driven diffusive
systems. We consider the broad class of models where particles can hop
between lanes but where the motion within each lane is not influenced
by other lanes. {(For examples of the converse case in which particles
  cannot hop between lanes but where the motion {within} each lane
  depends on the occupancy of the other lane,
  see~\cite{MRFF10,PP01,PS2010}.)}  The dynamics in each of the
  lanes can have different {characteristics}, for example, asymmetric
  exclusion dynamics in one lane and diffusive dynamics in the other.
We show that the phase diagram of such models can be constructed using
a new {\it stability analysis}. The stability analysis is shown to be
equivalent to an extremal current principle for two-lane systems
and both methods may be used according to convenience. We illustrate
the usefulness of the method for several systems and show that
additional lanes generically lead to the appearance of new phases in
the system. Finally, we argue that when the motion on each lane is
influenced by the other lane the extremal current principle does
not generically hold.

The paper is organized as follows. In section~(\ref{sec:models}) we
define the class of models that we shall study, and their mean-field
dynamics. In section~(\ref{sec:StabAn}) we present the stability
analysis method for deriving phase diagrams of two-lane models. We
illustrate this approach for a simple case in
section~\eqref{sec:tasep1l} and then consider richer models in
section~\eqref{sec:othermodels}. In section~(\ref{app:WI}), we show an
example of a more general class of systems where both our stability
analysis and the extremal current principle may not apply. Finally, in
section~\eqref{sec:conc} we conclude.

\section{Model definition and mean-field equations}    
\label{sec:models}
\subsection{The class of models}
Throughout the paper we consider a class of driven diffusive systems
composed of two one-dimensional lattices (or `lanes').  Each lane
has $L$ sites labelled $i\in[1,L]$ along which particles can hop.  In
addition, particles can hop between different lanes, the dynamics of
which are thus effectively coupled.  Each lane is connected at its
ends to reservoirs which have specified particle densities.

\begin{figure}[ht]
  \begin{center}
    \includegraphics{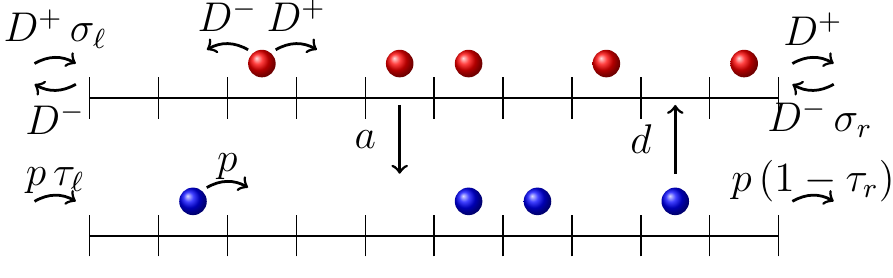}
  \end{center}
  \caption{Example of two coupled lattice gases in contact with
    reservoirs -- Model I. The occupancies of sites $i$ of the lower and upper
    lattices are denoted by $\tau_i$ and $\sigma_{i}$, respectively. The
    bottom lattice has at most one particle per site and the particles
    can only jump to the right, at rate $p$ (totally asymmetric exclusion
    dynamics). On the upper lattice, particles hop freely to the
    right with rate $D^+$ and to the left with rate $D^-$ (biased diffusive
    dynamics). Particles on the bottom lattice hop with rate $d$ to the upper lattice
    whereas those from the upper lattice hop with rate $a$ to empty
    sites of the lower lattice. Both lattices are coupled to
    reservoirs with densities $\tau_\ell,\sigma_\ell$ and
    $\tau_r,\sigma_r$ at their left and right ends, respectively.}
    \label{fig:TASEP-DIFF}
  \end{figure}

We will generally  make the following assumptions:
\begin{enumerate}
\item\label{local} The particles hop locally, i.e.~particles can
only hop from a site $i$ to a neighbouring site in that
lane, or to the same site $i$ on another lane.
(See for example figure~\ref{fig:TASEP-DIFF}.) 

\item\label{nonint}The hopping rates of the particles within a lane
  depend only on the occupancies of their site and neighbouring sites
  {\em within} their lane (not on the occupancies of sites in the
  neighbouring lane).

\item\label{stab} The average transverse flux of particles from one
lane to the other increases with the occupancy of the departing lane
and decreases with that of the arriving one.

\item\label{equil} The reservoir densities are equilibrated, i.e.~the
  densities are always chosen such that they produce zero transverse
  flux.

\item\label{trnas} The dynamics in the bulk are translationally invariant.
\end{enumerate}

We note that (\ref{local}) and (\ref{stab}) are physically reasonable
assumptions, whereas (\ref{nonint}), (\ref{equil}) and (\ref{trnas})
are simplifying assumptions.  {In~\ref{app:A2}} we discuss the
effect of relaxing assumption (\ref{equil}) whereas a full study of
the effects of relaxing (\ref{nonint}) (i.e.~allowing direct coupling
between the flows in different lanes) is beyond the scope of the
present paper. We do, however, argue {in section~(\ref{app:WI}}) that under
generic conditions the methods developed in this paper may fail when
this occurs {and show a particular example where this happens}.

For a concrete example of a model {in the class described above}, see
figure~\ref{fig:TASEP-DIFF}, which illustrates a simple two-lane
system.  The dynamics of the bottom lane is that of a totally
asymmetric simple exclusion process (TASEP). Namely, particles hop
only to the right, with rate $p$, provided the arrival site is
empty. In the upper lane, there is no exclusion and particles hop to
the left and to the right with rates $D^-$ and $D^+$ respectively,
i.e.~the particles perform biased diffusive motion.  A particle on
site $i$ of the bottom lane can also hop to site $i$ of the upper lane
with rate $d$. A particle on site $i$ of the upper lane can hop to site $i$ on the lower
lane with rate $a$ provided that the arrival site is empty.  Note that there are no explicit interactions
between particles on different lanes---the hopping rate of the
particles along each lane does not depend on the occupancies of the
other lane.

We refer to the model presented in figure~\ref{fig:TASEP-DIFF} as
\emph{Model I}.  We shall return to it throughout and use it as a
template with which to illustrate our approach, before considering
applications to other models towards the end of the paper.

\subsection{Mean-field equation}

To analyze the phase diagram of driven diffusive systems, it is often
sufficient to consider a simple mean-field approximation, wherein one
replaces the $n$-point correlation function by the product of $n$
one-point correlation functions. For the average occupancies, one then
ends up with equations of the form
\begin{equation}
  \dot \tau_i=-{\cal J}^\tau_i+{\cal J}^\tau_{i-1}+K_i\;; \qquad\dot
    \sigma_i=-{\cal J}^\sigma_i+{\cal J}^\sigma_{i-1}-K_i,
\label{mf}
  \end{equation}
  where $\tau_i$ and $\sigma_i$ are the average occupancies of the
  lower and upper lanes at site $i$; ${\cal J}^\tau_i$, ${\cal
    J}^\sigma_i$ are the mean-field approximations of the current
  between sites $i$ and $i+1$ along the lower and upper lanes,
  respectively. On the boundaries, the occupancies of both lanes equal
  those of the reservoirs, $\tau_l,\sigma_l$ and $\tau_r,\sigma_r$
  (see figure 1). $K_i$ represents the mean-field value of the 
  transverse currents between the two lanes. 

\if{Our assumption (\ref{stab}) about the hopping
  rates between the two lanes made in the previous section amounts
  to
\begin{equation}
  \partial_{\tau_i} K_i < 0 < \partial_{\sigma_i} K_i \;.
  \label{eqn:Keq}
\end{equation}
That is, the transverse current $K_i$ increases and decreases with the
occupancies of the departing and arriving sites, respectively.  }\fi

For Model I (see figure 1) the currents are given by
  \begin{eqnarray}
    {\cal J}^{\tau}_i=p\tau_i (1-\tau_{i+1}) \nonumber \\
    {\cal J}^\sigma_i=D^+\sigma_{i}-D^-\sigma_{i+1} \nonumber \\
    K_i=-d \tau_i+a\sigma_i(1-\tau_i).
\label{eqn:disCur}
  \end{eqnarray}

Although our approach could be carried out at the level of
spatially discrete equations (\ref{mf}),  it is
convenient to take the continuum limit as detailed below.

\subsection{Continuum limit}
\label{sec:mfcteqn}
We take the continuum limit and keep terms up to second order in
gradients. The mean-field equations are then given by
\begin{eqnarray}
  \label{eq:MFeqcont}
  \dot \tau &=& - \partial_x [J_\tau- D_\tau \partial_x \tau] +K(\tau,\sigma) \\ \dot \sigma &=& -
  \partial_x[ J_\sigma - D_\sigma \partial_x \sigma] -K(\tau,\sigma).\nonumber
\end{eqnarray}
Here we have explicitly separated the
mean field currents ${\cal J}_\tau$ and ${\cal J}_\sigma$
into their advective parts, $J_\tau$ and $J_\sigma$,
and their diffusive parts, $-D_\tau \partial_x \tau$ and $-D_\sigma \partial_x \sigma$. The advective currents
$J_\tau$ and $J_\sigma$ are non-zero only when the dynamics on the
corresponding lane are driven or equivalently when the hopping rates
are asymmetric.
Equations (\ref{eq:MFeqcont})  have to be complemented with boundary
conditions on the values of $\sigma$ and $\tau$ which are imposed by
the reservoirs:
\begin{equation}
\qquad \tau(0)=\tau_l\qquad\tau(L)=\tau_r  \qquad \sigma(0)=\sigma_l\qquad\sigma(L)=\sigma_r.
\label{eq:bc}
\end{equation}
Following assumption (\ref{equil}) we take the boundary densities to
be equilibrated, that is
\begin{equation}
K(\tau_\ell,\sigma_\ell)=K(\tau_r,\sigma_r)=0\;.
\label{bequil}
\end{equation}
We discuss the general case where (\ref{bequil}) does not hold
in \ref{app:A2}.

Our assumption (\ref{stab}) about the hopping
rates between the two lanes made in the previous section amounts
  to
\begin{equation}
  \partial_{\tau} K < 0 < \partial_{\sigma} K \;.
  \label{eqn:Keq}
\end{equation}
That is, the transverse current $K$ increases and decreases with the
occupancies of the departing and arriving sites, respectively.  

In the case of Model I (figure~\ref{fig:TASEP-DIFF}), 
it is straightforward to deduce 
the currents from Eq.~\ref{eqn:disCur} 
\begin{eqnarray}
  \label{eqn:currentex1}
  J_\tau &=& p \tau (1-\tau) \\
  J_\sigma&=&v \sigma \nonumber\\
  K&=&a \sigma (1-\tau) - d \tau \;.\nonumber
\end{eqnarray}
Here $v= D^+-D^-$, $D_\tau=p/2$ and $D_\sigma =(D^++D^-)/2$ and we set the lattice constant to unity.  Thus,
when $D^+=D^-$, $v=0$ and the model reduces to that of \cite{KL03}.
Also note that $ J_\tau = J_\tau(\tau)$ and $J_\sigma =
J_\sigma(\sigma)$, which is consistent with assumption (\ref{nonint}),
and that this choice of $K$ satisfies condition (\ref{eqn:Keq}).

\section{Stability analysis}
\label{sec:StabAn}
Constant (i.e.~flat) profiles $\tau(x)=\tau_0$ and
 $\sigma(x)=\sigma_0$ are solutions of the bulk equations~\eqref{eq:MFeqcont}
 as long as they satisfy
\begin{equation}
  \label{eqn:k0}
  K(\tau_0,\sigma_0)=0.
\end{equation}
We refer to such constant solutions as equilibrated plateaux. Because
of our assumption \eqref{eqn:Keq} on the dependence of $K$ on $\tau$
and $\sigma$, the relation $K(\tau_0,\sigma_0)=0$ implies that
$\sigma_0$ is an increasing function of $\tau_0$. Since the hopping
rate within each lane does not depend on the occupancy of the other
lane -- see assumption (ii) -- one can show that such solutions are
{\em dynamically stable}. The details are left to \ref{app:dynstab}
where it is {shown} that perturbations about the plateaux $\delta
\tau(x,t)=\tau(x,t)-\tau_0$ and $\delta
\sigma(x,t)=\sigma(x,t)-\sigma_0$ decay exponentially in time. On the
other hand, we provide in section~(\ref{app:WI}) an example of a more
general two-lane system where assumption (ii) does not hold and where
equilibrated plateaux are not necessarily dynamically stable.

Let us now consider the steady-state version of the mean-field
equations~\eqref{eq:MFeqcont}
\begin{eqnarray}
  \partial_x(D_\tau \partial_x \tau) = \partial_x J_\tau(\tau) - K(\tau,\sigma)\label{eqn:2ltau}\\
  \partial_x (D_\sigma \partial_x \sigma)  = \partial_x J_\sigma(\sigma) + K(\tau,\sigma)\label{eqn:2lsigma}
\end{eqnarray}
along with the boundary conditions (\ref{eq:bc}). Generally, constant
profiles will not satisfy the boundary conditions~\eqref{eq:bc} on
both sides since the left and right reservoirs may impose different
densities.  It is thus natural to ask whether one can construct
steady-state profiles by matching two sets of plateaux that separately
satisfy the left and right boundary conditions. As we now show,
answering this question requires the study of the {\em spatial}
stability of perturbations around the plateaux. We refer to
perturbations as diverging if they grow as $x$ increases and
converging if they decrease as $x$ increases and we refer to the
corresponding plateaux as stable or unstable. It is important to note
that from now on, any reference to {\em stability} or {\em
  instability} has to be understood with reference to the spatial
dependence of a profile on $x$, since plateaux are always {\em
  dynamically} stable.

\begin{figure}[ht]
  \begin{center}
    \includegraphics{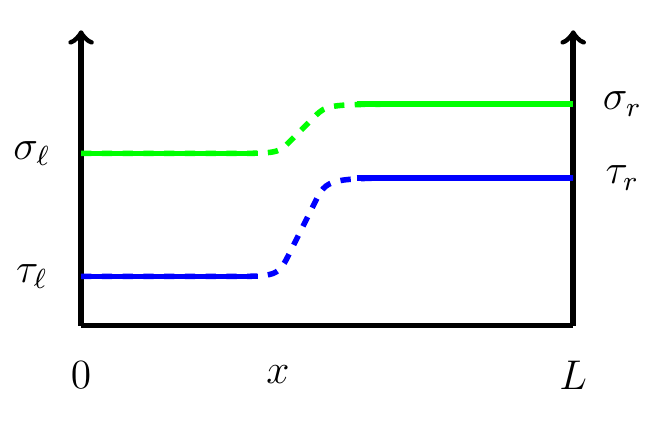}
  \end{center}
  \caption{Plateaux imposed by the
    `equilibrated' densities of left and right reservoirs are simple
    steady-state solutions of the mean-field equations at both ends of
    the system. It remains to be known whether the connecting shock
    is an acceptable steady-state solution.}
  \label{figure:gluing}
\end{figure}

In the example presented in figure \ref{figure:gluing}, for instance,
one needs the profiles to diverge away from the left plateaux and
converge into the right ones as $x$ increases. Since $\sigma$ is an
increasing function of $\tau$ in equilibrated plateaux, the
perturbations for $\tau$ and $\sigma$ away from the boundary plateaux
must be of the same sign~\footnote{Note that this assumes that the
  profiles connecting the plateaux are monotonous.}.

To check that this indeed can happen we carry out a stability analysis
of Eq. \eqref{eqn:2ltau} and \eqref{eqn:2lsigma}. Consider a
small perturbation $(\epsilon(x),\eta(x))$ around equilibrated plateaux so
that $\tau(x)=\tau_0+\epsilon(x)$ and $\sigma(x)=\sigma_0+\eta(x)$
where $\tau_0$ and $\sigma_0$ {satisfy} $K(\tau_0,\sigma_0)=0$.
Expanding the steady-state mean-field equations to first order in
$\epsilon,\, \eta$ gives
\begin{eqnarray}
  \label{eqn:linsyst}
  D_\tau \partial_{x}^2 \epsilon = (\partial_\tau J_\tau)\partial_x \epsilon
 - (\partial_\sigma K) \eta - (\partial_\tau K) \epsilon\\
  D_\sigma \partial_{x}^2 \eta = (\partial_\sigma J_\sigma)
  \partial_x \eta+ (\partial_\sigma K) \eta + (\partial_\tau K) \epsilon.\nonumber
\end{eqnarray}
Here the derivatives with respect to $\tau$ and $\sigma$ are taken at
$\tau_0$ and $\sigma_0$ respectively and similarly $D_\tau$ and
$D_\sigma$ are evaluated at $\tau_0$ and $\sigma_0$
respectively. Looking for solutions of this system of two second order
linear differential equations of the form
\begin{equation}
(\epsilon(x),\eta(x))=\exp(\lambda x) (\epsilon_0,\eta_0),
\label{pert}
\end{equation}
equations \eqref{eqn:linsyst} become
 \begin{equation}
   \label{eqn:linmat}
  0 = \smallmatrix{
    \lambda^2 D_\tau - \lambda \partial_\tau J_\tau +\partial_\tau K & \partial_\sigma K\cr
    -\partial_\tau K &      \lambda^2 D_\sigma-\lambda \partial_\sigma J_\sigma - \partial_\sigma K\cr}  \smallmatrix{\epsilon_0\cr\eta_0\cr}.
\end{equation}
For non-trivial solutions to exist, one needs the determinant of the
matrix in (\ref{eqn:linmat}) to vanish, that is
\begin{equation}
  \lambda\, \chi(\lambda)=0\label{eqn:poleqx}
\end{equation}
where  
\begin{eqnarray}
\chi(\lambda)&=& D_\tau D_\sigma \lambda^3 -\lambda^2(D_\tau \partial_\sigma J_\sigma+D_\sigma
  \partial_\tau J_\tau) +\lambda( \partial_\sigma J_\sigma\, \partial_\tau J_\tau  \label{eqn:bouh}\\ 
  && +D_\sigma
  \partial_\tau K-D_\tau \partial_\sigma K)\nonumber +\partial_\sigma  K\, \partial_\tau J_\tau
-\partial_\tau K\, \partial_\sigma J_\sigma \; .
\end{eqnarray}
The spatial stability of the plateaux is then determined by the four
solutions of the polynomial equation \eqref{eqn:poleqx}. 
In  \ref{sec:appendixa} we prove that the four solutions,
which we denote $\lambda_0$, $\lambda_1$, $\lambda_2$, $\lambda_3$,
are real and  have the following properties:

\begin{itemize}
\item There is a  trivial solution $\lambda_0=0$ which
corresponds to shifting the values of the two plateaux while keeping
$K=0$, i.e.~$(\epsilon_0,\eta_0) \propto (-\partial_\sigma K,
\partial_\tau K)$; 
\item There is exactly one additional solution such that $\epsilon_0
  \eta_0 >0$.  The corresponding root $\lambda_1$ is the only root
  that may change sign as the parameters {$(\tau_0,\sigma_0)$} are
  varied and its sign is given by the sign of $\chi(0)$.
\item There are two roots $\lambda _2 <0$ and $\lambda_3>0$
with $\epsilon_0 \eta_0 <0$.  

\item The root $\lambda_1$ is the only
root of~\eqref{eqn:poleqx} relevant for the derivation of the phase
diagram:
it controls the convergence  of
perturbations away from constant  densities.

\item The roots $\lambda_2$ and $\lambda_3$ only play a role for the boundary layers
connecting the plateaux to {\em non-equilibrated}
reservoirs (i.e.~with $K\neq 0$ at the boundary) and not for
the construction of the phase diagram (see \ref{app:A2}). 
\end{itemize}

The perturbations $(\epsilon,\eta)$ associated to $\lambda_1>0$ thus
grow as $x$ increases (a diverging perturbation), while those associated to $\lambda_1<0$
decrease (a converging perturbation). From equation~\eqref{eqn:bouh} it readily follows that the
change of spatial stability occurs when
\begin{equation}
  \label{eqn:stabeq}
  \chi(0)=\partial_\sigma K(\tau_0,\sigma_0)\partial_\tau J_\tau(\tau_0)-\partial_\tau K(\tau_0,\sigma_0) \partial_\sigma J_\sigma(\sigma_0)=0.
\end{equation}
Thus the perturbations $(\epsilon,\eta)$ grow or decrease with $x$ when $\chi(0)$ is negative or
positive, respectively (see \ref{app:A1}).

The method of constructing the phase diagram is then to determine the
sign of $\chi(0)$ for each set of equilibrated plateaux (imposed by
the boundaries) and deduce whether the plateaux can be connected by
appropriate perturbations. For example, the scenario presented in
figure 2 requires a diverging perturbation from the left plateau and a
converging perturbation at the right plateau.  This can only take
place if $\chi(0)>0$ for the left plateau densities $\tau_l$,
$\sigma_l$ and $\chi(0)<0$ for the right ones $\tau_r$, $\sigma_r$. As
we illustrate below, this line of reasoning provides a straightforward
method to construct the phase diagram of multilane systems.

Before turning to derive the phase diagram for several examples, let us show
that  $\chi(0)$ is simply related to the derivative of the total
current. In equilibrated plateaux, $\sigma_0$ and $\tau_0$ are related
through $K(\tau_0,\sigma_0)=0$. The advective part of the total
current, defined as
\begin{equation}
J_{\rm tot}=J_\tau+J_\sigma\;,
\end{equation}
may  thus be expressed as a function of
$\tau_0$ whose total derivative is given by
\begin{equation}
  \frac{{\rm d} }{{\rm d}\tau_0} J_{\rm tot}\big(\tau_0,\sigma_0(\tau_0)\big)= \partial_{\tau_0} J_{\rm
  tot}+\partial_{\sigma_0} J_{\rm tot} \frac{\partial\sigma_0}{{\partial
  \tau_0}} =\partial_{\tau_0} J_{\tau_0}- \partial_{\sigma_0} J_{\sigma_0}
\frac{\partial_{\tau_0} K}{\partial_{\sigma_0} K} .
\end{equation}
Equation~\eqref{eqn:stabeq} then implies
\begin{equation}
\label{eqn:chiisJ}  \chi(0)={\partial_{\sigma_0} K} \  \frac{{\rm d} J_{\rm tot}}{{\rm d}\tau_0}.
\end{equation}
Since under our assumptions (see Eq. (\ref{eqn:Keq})) $\partial_{\sigma_0} K >0$, the sign of $\chi(0)$ is the same as that of $\frac{{\rm d}
  J_{\rm tot}}{{\rm d}\tau_0}$. Therefore, the perturbation connecting different
plateaux changes stability at an extremum of the advective part of the
total current $J_{\rm tot}$. It is converging when $\frac{{\rm d}J_{\rm
    tot}}{{\rm d}\tau_0}<0$ and diverging when $\frac{{\rm d}J_{\rm
    tot}}{{\rm d}\tau_0}>0$.

{Note that the case $\frac{{\rm d}J_{\rm tot}}{{\rm d}\tau_0}=0$
  requires a special treatment. Maximal current plateaux (where
  $\frac{{\rm d}^2J_{\rm tot}}{{\rm d}\tau_0^2}<0$) are such that
  negative perturbations are diverging whereas positive perturbations
  are converging, so that profiles connected to these plateaux are
  always decreasing. Conversely, minimal current phases correspond to
  increasing profiles, i.e. $\tau(x)$ and $\sigma(x)$ increasing with
  $x$.}

\section{Construction of the phase diagram -- a simple case}
\label{sec:tasep1l}

To demonstrate how the stability analysis described above can be used
to construct phase diagrams of different microscopic models we first
study Model I when $D^+=D^-$ and thus $v=0$ (symmetric
diffusion). Later we also consider the more general case where $v \neq
0$.

\begin{figure}[ht]
  \begin{center}
\raisebox{1cm}{\includegraphics{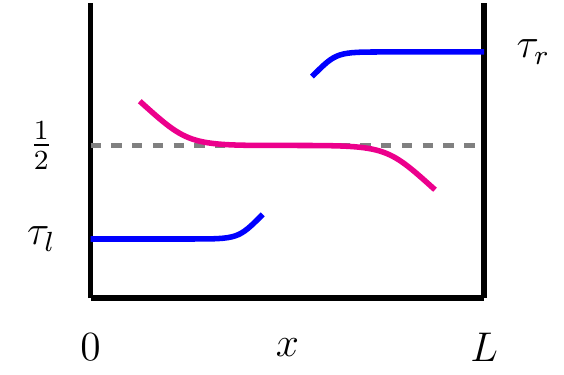}}
    \hspace{1cm}\includegraphics{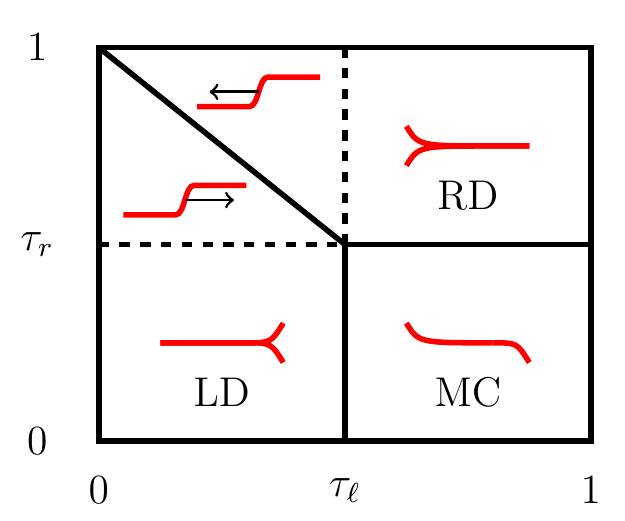}
  \end{center}
  \caption{{\bf Left}: The left and right reservoirs impose boundary
    densities $\tau_\ell$ and $\tau_r$. Various possible steady
    profiles are presented, together with their stability. {\bf
      Right}: Phase diagram constrained by the stability of the
    equilibrated plateaux presented on the left panel.}
  \label{BLAH}
\end{figure}

Using equations~\eqref{eqn:currentex1} and \eqref{eqn:chiisJ}, we see
that the stability of the equilibrated plateau is set by the sign
of
\begin{equation}
  \frac{{\rm d}J_{\rm tot}}{{\rm d}\tau_0}=p (1-2\tau_0) \label{eqn:stupidcase}.
\end{equation}
This example is particularly simple because the advective part of the
total current does not depend on $\sigma_0$. Equation
\eqref{eqn:stupidcase} implies that profiles with $\tau_0< 1/2$ are
diverging (growing with $x$) whereas profiles with $\tau_0 > 1/2$ are
converging (decreasing with $x$). The case $\tau_0=1/2$ is marginal:
negative perturbations are diverging whereas and positive
perturbations are converging. {This is consistent with our claim,
  in the previous section, that maximal profiles correspond to
  $\tau(x)$ and $\sigma(x)$ decreasing with $x$.} The different
profiles are illustrated in figure~\ref{BLAH}.

The construction of the phase diagram is now straightforward. To do this we 
consider all possible matchings, four in total, between the different stability regions for
left and right boundary plateaux.

\begin{itemize}
\item When $\tau_\ell > 1/2$ and $\tau_r < 1/2$, the only profile
  consistent with the stability analysis is flat at $\tau(x)=1/2$
  connecting to the boundaries at the ends of the system. This
  corresponds to the maximal current phase {(MC phase)}. 
\item When $\tau_\ell < 1/2$ and $\tau_r < 1/2$, the only profile
  allowed by the stability analysis is a flat plateau with
  $\tau(x)=\tau_\ell$ and a short boundary layer connecting to the
  right boundary density {(LD phase)}.
\item When $\tau_\ell > 1/2$ and $\tau_r > 1/2$, similarly to the
  previous case, we have $\tau(x)=\tau_r$ and a short boundary layer
  connecting to the left boundary density {(RD phase)}.
\item When $\tau_\ell < 1/2$ and $\tau_r > 1/2$, the stability
  analysis is consistent with a shock profile {connecting
  $\tau(x)=\tau_\ell$ to $\tau(x)=\tau_r$}. The velocity of the shock,
  $v_s=(J_\tau(\tau_\ell)-J_\tau(\tau_r))/(\tau_r-\tau_\ell)$ is given
  by mass conservation. Therefore, the shock is steady for
  $\tau_r=1-\tau_\ell$ and is localized on the right or the left
  boundary for $\tau_r > 1- \tau_\ell$ and $\tau_r < 1 - \tau_\ell$
  respectively.
\end{itemize}

The corresponding phase diagram is drawn in figure~\ref{BLAH} and is
consistent with the results presented in~\cite{KL03}. 
This phase diagram reduces to that of the TASEP because of the
lack of bias on the diffusive lane. 
In effect the diffusive lane adjusts its density
to ensure that there is zero transverse flux  between the 
lanes, therefore the phase diagram is determined by that of the driven
lane i.e.~the TASEP.

\section{Two-lane models with non-trivial phase diagrams}
\label{sec:othermodels}
In this section we consider more general two-lane systems
which exhibit non-trivial phase diagrams.  In these cases the phase
diagrams are much richer than the TASEP coupled to a symmetric
diffusive lane considered in the previous section, and we analyse in
detail the phase diagrams of several such examples.

\subsection{TASEP and biased diffusion}
In this subsection we consider a TASEP coupled to a {\em
    biased} diffusive lane which corresponds to $v>0$ in Model I. Such
  a model is relevant to a biophysical situation where molecular
  motors diffuse asymmetrically after detaching from protofilaments,
  as for instance in models of cooperative extraction of membrane
  tubes~\cite{TEK09}.

Let us first show that in the presence of a bias in the diffusive
lane, the phase diagram of Model I can become more complex and exhibit
several new phases. As explained in section~\ref{sec:StabAn}, the
phase diagram can be deduced from $\dd J_{\rm tot}/\dd \tau_0$. From the
continuous mean-field equations \eqref{eq:MFeqcont} and
\eqref{eqn:currentex1}, one obtains
\begin{equation}
  J_{\rm tot}=p \tau(1-\tau) + v \sigma.
\end{equation}
{For equilibrated plateaux}, $K=0$ imposes
\begin{equation}
  \sigma_0=\frac da \frac{\tau_0}{1-\tau_0},
\end{equation}
so that
\begin{equation}
  J_{\rm tot}(\tau_0)=\frac {vd}a \frac{\tau_0}{1-\tau_0} + p\tau_0(1-\tau_0).
\end{equation}
The  values of $\tau_0$ where the stability
changes are given by ~$\dd J_{\rm tot}(\tau_0)/\dd \tau_0=0$ which reads
\begin{equation}
  (1-2\tau_0) (\tau_0-1)^2=-\frac{v d}{p a}.
\label{eqq:vdpa}
\end{equation}
Simple algebra shows that this equation always has an unphysical root
at $\tau_0>1$. Furthermore, one can distinguish two regimes, according
to the value of $vd/pa$, which exhibit distinct phase diagrams as we
now discuss.\\

\noindent{\bf Regime $v > p a/27d$}: in this case there are no
physical roots of equation (\ref{eqq:vdpa}). One finds that the 
plateaux are always diverging.  Since all profiles are diverging, there
is only one phase. One only observes a plateau with a density given by
the density at the left boundary and a small boundary layer that
connects to the right boundary. In this regime the advection
  in the diffusive lane is sufficiently strong that the left boundary
  always dominates.\\

\begin{figure}[ht]
  \begin{center}
\includegraphics{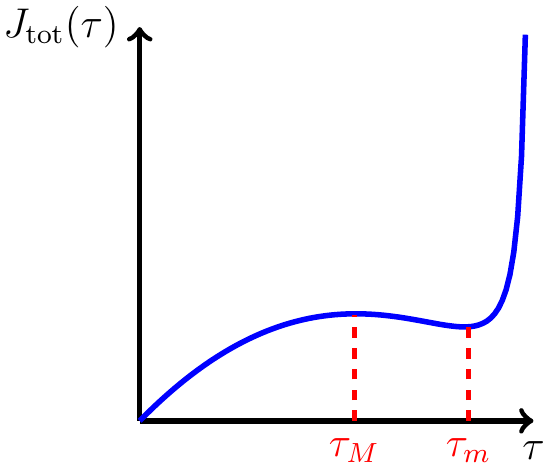}\hspace{1cm}    \includegraphics{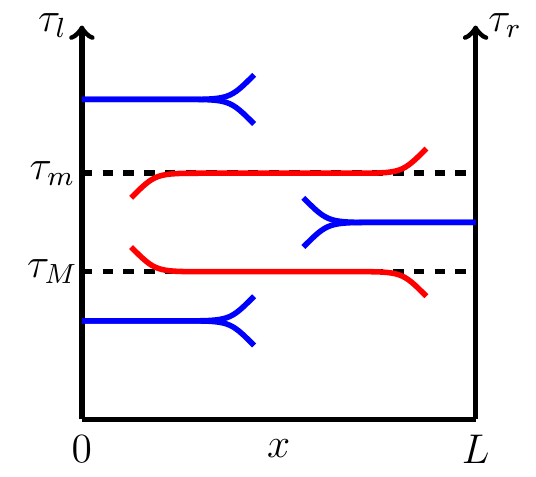}\\
    \includegraphics[width=.95\textwidth]{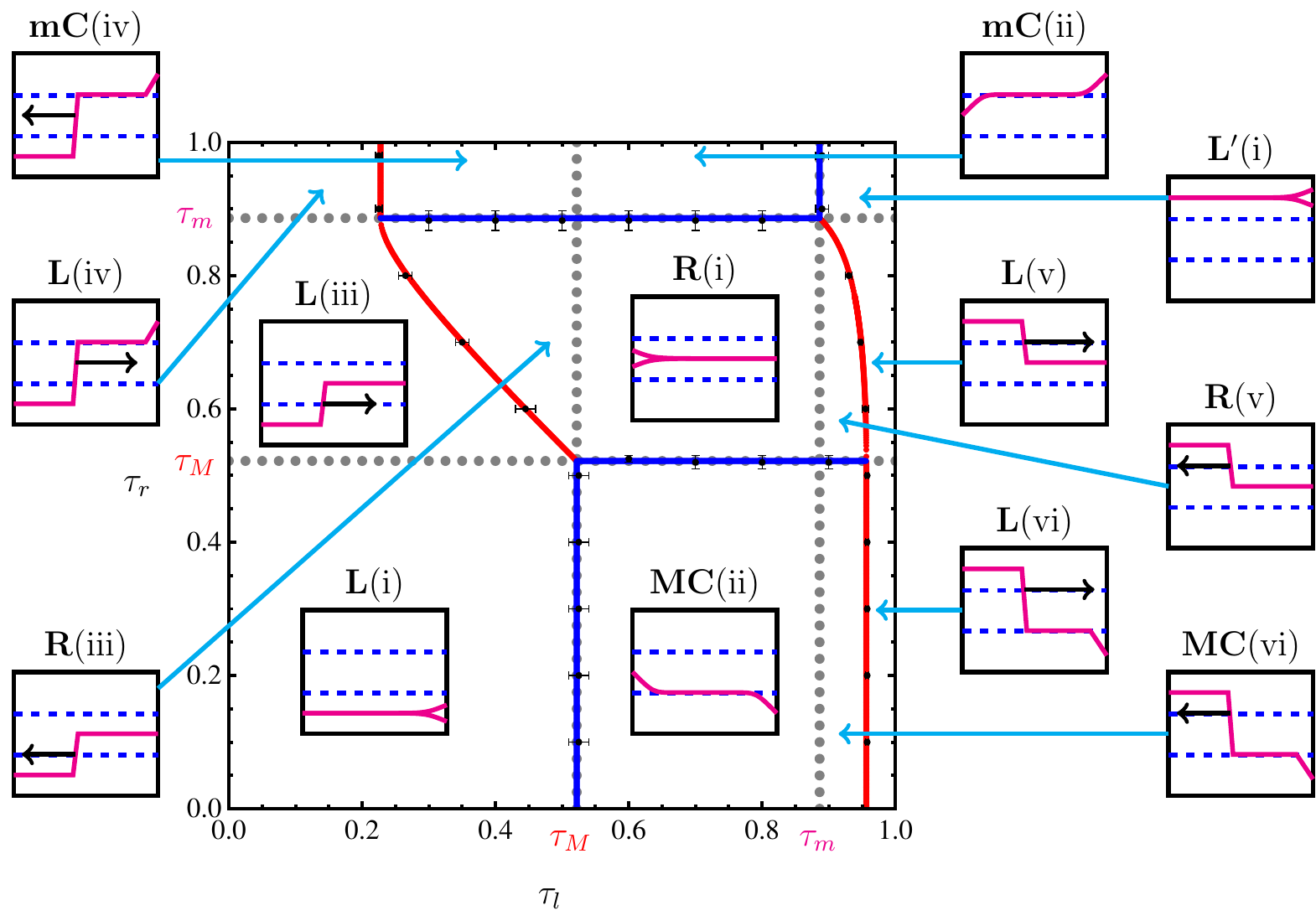}
  \end{center}
  \caption{{\bf Top-left}: Current-density relation for the TASEP
    coupled to a biased diffusive lane. {\bf Top-right} This figure
    represents a side view of steady density profiles that can be used
    to connect two reservoirs with densities $\tau_r$ and
    $\tau_l$. According to the stability analysis, in the steady
    states, one can get diverging profiles for $\tau \leq \tau^M$ or
    $\tau \geq \tau^m$ and converging profiles otherwise. Note that the extremal densities have been spread out for clarity and their values are just illustrative. {\bf Bottom:} Phase
    diagram of a TASEP coupled to a biased diffusive lane, for
    parameters $p=1,\,d=0.01,\,a=1,\,D=1,\,v=1$. The blue and red lines
    correspond to continuous and discontinuous phase transitions
    predicted by the theory while the black symbols were obtained from
    continuous time Monte-Carlo simulations. The grey dotted lines
    correspond to the local minimum and maximum of $J_{\rm
      tot}$. Insets illustrate the stability discussion of the
      main text and the consequent profiles observed in the steady
      state. $L$ and $R$ stand for left- and right-dominated phase
    while mC and MC stand for minimal and maximal current phase.}
  \label{figure:stability}  \label{fig:phaseD}
\end{figure}

\noindent{\bf Regime ${pa}/27d>v >0$}: in this case there are two
roots of (\ref{eqq:vdpa}), denoted $\tau_M$ and $\tau_m$, which lie
between 0 and 1.  Therefore the stability of the plateau changes
twice: it is diverging for $\tau_0$ close to 0 and 1 and is converging
in the intermediate region (see
figure~\ref{figure:stability}). {The smallest root $\tau_M$ is a
  local maximum of $J_{\rm tot}(\tau_0)$, so that profiles that are
  connected to the corresponding plateau are decreasing. On the
  contrary, the plateau at $\tau_m$ can only be connected to 
  increasing profiles (see figure~\ref{figure:stability}).}

We now illustrate how the phase diagram can
be constructed using figure~\ref{figure:stability}. Once again, we
look at possible matchings of the plateaux imposed by the boundary
conditions. Both $\tau_\ell$ and $\tau_r$ can be in the three
different regions separated by $\tau_M$ and $\tau_m$. 
As indicated in figure~\ref{fig:phaseD}, each of these
nine cases has to be analyzed separately and the result is five distinct phases
for any $d,a\neq 0$.  Two of the
phases disappear at $d=0$.
\begin{enumerate}
\item {\bf Regions $R(i)$, $L(i)$ and $L'(i)$}: if the left and right
  densities are in the same stability region (a total of 3 different
  possibilities) there is one plateau set by one boundary and a small
  boundary layer that connects to the other boundary. For diverging
  regions ($\tau_r,\tau_\ell<\tau_M$ and $\tau_r,\tau_\ell>\tau_m$)
  the left density controls the plateau whereas the right boundary
  dominates in the converging region
  ($\tau_M<\tau_r,\tau_\ell<\tau_m$).

\item {\bf Regions $mC(ii)$ and $MC(ii)$}: when
  $\tau_M<\tau_\ell<\tau_m$ and $\tau_r<\tau_M$, the stability implies
  a plateau at $\tau_M$. Since this value corresponds to a local
  maximum of the current, we conclude that this region belongs to a
  maximal current phase. Similarly, when $\tau_M<\tau_\ell<\tau_m$ and
  $\tau_r>\tau_m$, the stability implies a plateau at $\tau_m$. This
  region therefore belongs to a minimal current phase.

\item {\bf Regions $L(iii)$ and $R(iii)$}: when $\tau_\ell<\tau_M$ and
  $\tau_M< \tau_r <\tau_m$, the stability analysis predicts that the
  corresponding plateaux can be connected by an upward shock, from
  low to high density plateau. Since the total density
  $\rho=\tau+\sigma$ is locally conserved, the shock velocity is given
  by
\begin{equation} 
  \label{eqn:vshock}
  v_s=[J_{\rm tot}(\tau_r)-J_{\rm tot}(\tau_\ell)]/(\rho_r-\rho_\ell).
\end{equation}
Here, $\rho_r>\rho_\ell$ and the plateau associated with the smallest
current extends whereas the other recedes. The limiting case $J_{\rm
  tot}(\tau_r)=J_{\rm tot}(\tau_\ell)$ defines a boundary line
$\tau_\ell(\tau_r)$ {separating a `right-dominated' phase $R(iii)$ from
a `left-dominated' one $L(iii)$}, corresponding to a discontinuous
phase transition. On this line, a shock in the profile can in
principle be observed.

\item {\bf Regions $L(iv)$ and $mC(iv)$}: when $\tau_\ell<\tau_M$ and
  $\tau_r>\tau_m$, the stability analysis predicts that a plateau of
  density $\tau_\ell$ can be connected by a shock to a plateau of
  density $\tau_m$ which is then connected to the right density
  through a small boundary layer. This phase can be either
  `left-dominated' or in a minimal current phase depending on the sign
  of the shock velocity. Since the shock is upward, the plateau
  with the lowest current dominates. A shock can in principle be
  observed on the transition line where $J_{\rm tot}(\tau_\ell)=J_{\rm
    tot}(\tau_m)$.

\item {\bf Regions $L(v)$ and $R(v)$}: when $\tau_\ell > \tau_m$ and
  $\tau_M<\tau_r<\tau_m$ the stability analysis allows a downward
  shock to form in the system. Similar to the arguments regarding
  shocks presented above since $\rho_\ell > \rho_r$ the plateau
  associated with the larger current controls the system. The phase
  transition line is given by $J_{\rm tot}(\tau_r)=J_{\rm
    tot}(\tau_\ell)$.

\item {\bf Regions $L(vi)$ and $MC(vi)$}: when $\tau_\ell > \tau_m$
  and $\tau_r<\tau_M$ the stability analysis allows a downward shock
  to form in the system {from $\tau(x)=\tau_\ell$ to a lower density given
  by $\tau_M$. The latter is then connected to the right
    boundary} with a short boundary layer. The scenario is depicted in
  figure \ref{fig:phaseD}. Again the larger current controls the
  system and there is a phase transition when $J_{\rm
    tot}(\tau_M)=J_{\rm tot}(\tau_\ell)$.

\end{enumerate}

Note that the contiguous regions $L(i)$, $L(iii)$, $L(iv)$ are all
controlled by the density at the left end of the system and therefore
constitute a single, low-density, phase, with plateaux density smaller
than $\tau_M$. Similarly, the three contiguous phases $L(v)$, $L(vi)$
and $L'(i)$ form a single high-density phase, in which the plateau
density is larger than $\tau_m$.  Then, each of the groups of regions
[$R(i)$, $R(iii)$, $R(v)$], [$MC(ii)$,$MC(vi)$], [$mC(iv)$, $mC(ii)$]
defines a different phase.

To summarize this subsection, we have constructed the phase diagram
using a stability analysis. The resulting phase diagram has five
distinct phases. In figure \ref{fig:phaseD} the results of the
analysis are compared to continuous time Monte-Carlo simulations and
shown to agree very well.

{Note that we show in~\ref{sec:appMSA} that the above stability
  analysis leads to the same phase diagram as a generalization of the
  extremal current principle to the two-lane systems. For {\em
    increasing} profiles ($\tau_\ell \leq \tau_r$) the steady plateaux
  are such that $J_{\rm tot}(\tau)$ realizes its {\em minimum} for
  $\tau \in [\tau_\ell,\tau_r]$. Conversely, for {\em decreasing}
  profiles ($\tau_\ell > \tau_r$), the steady profiles are such that
  $J_{\rm tot}(\tau)$ realizes its {\em maximum} for $\tau \in
  [\tau_r,\tau_\ell]$. Which of the two methods one chooses to use is
  thus a matter of convenience.}

Next, we derive the phase diagram for two other two-lane models and
compare the results to continuous time Monte-Carlo simulations. The
derivation follows very similar lines to that presented above and we
thus only outline the main differences.

\subsection{Two coupled TASEPs with advection in the same direction}

In this subsection we consider two coupled TASEPS
with advection in the same direction but of different strength in the two lanes.

For two coupled TASEPs with equilibrated boundaries, Harris and
Stinchcombe have shown that if the extremal current principle applies,
then the phase diagram should be the same as that of a single-lane
TASEP with next-nearest neighbours interactions~\cite{HS05}. They
backed up this assumption by simulating the profiles for several
points in the phase diagram. Here we derive the phase diagram using
the stability argument. Our results agree with those found
in~\cite{HS05} and with continuous time Monte-Carlo simulations (see
figure~\ref{fig:2TASEPFSPD}).

\begin{figure}[ht]
  \begin{center}
    \includegraphics{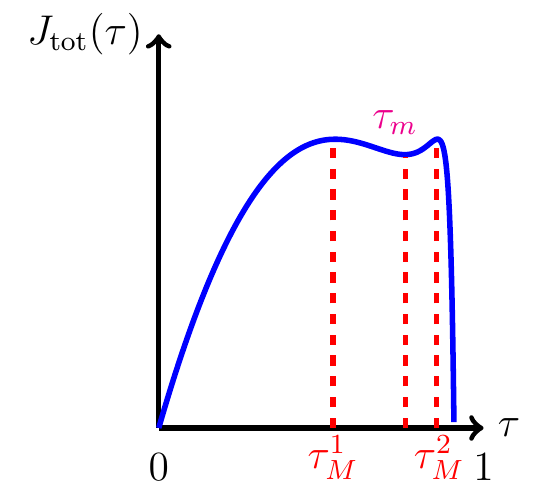}\hspace{2cm} \includegraphics{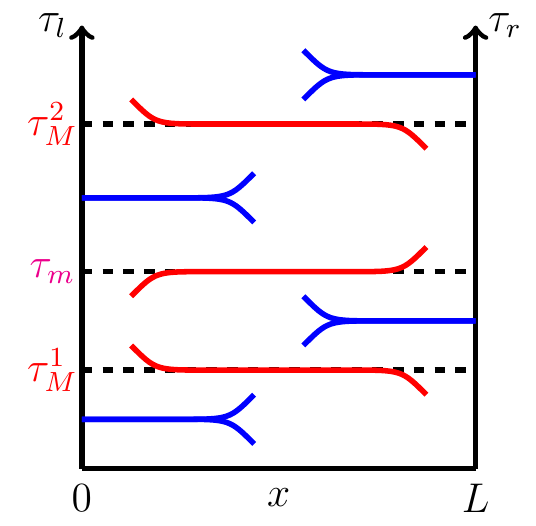}
  \end{center}
  \caption{{\bf Left:} Current density relation for two coupled
    TASEPs with a bias in the same direction. {\bf Right:} Sketch
    of the stability of various steady-state profiles for two coupled
    TASEPs with a bias in the same direction. Note that the extremal densities have been spread out for
    clarity and their values are just illustrative.}
  \label{fig:2TASEPFSCP}
\end{figure}

\begin{figure}[ht]
  \begin{center}
      \includegraphics{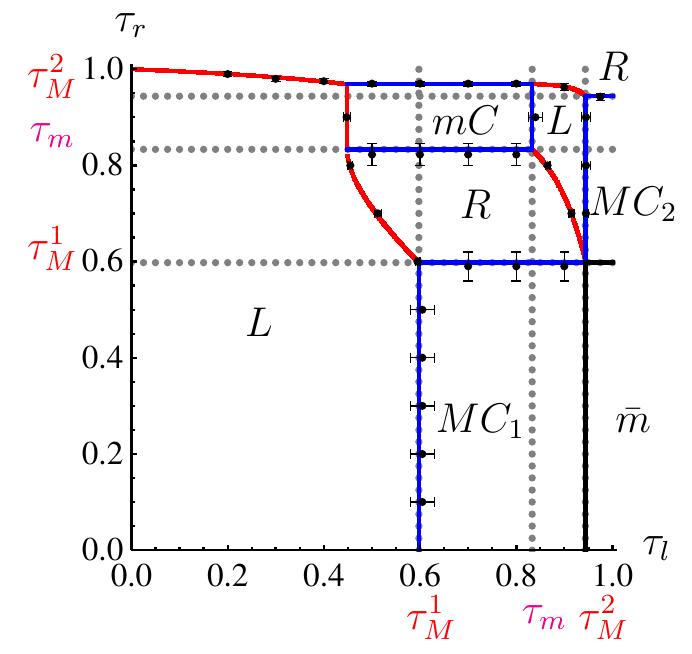}
  \end{center}
  \caption{Phase diagram of two forward TASEPs with
    $p=q=1,\,d=0.04,\,a=1$. {The blue and red lines correspond to the
    continuous and discontinous phase transitions predicted by the
    stability analysis. The dotted grey line indicate the extrema of
    the current and the black dots are the results of the continuous time
    Monte-Carlo simulations.}}
  \label{fig:2TASEPFSPD}
\end{figure}

The model we study in this subsection is defined as follows. We
consider two lanes, each consisting of a one-dimensional lattice with
$L$ sites. On the lower lane particles hop to a nearest neighbour empty
site to their right with rate $p$. On the upper lane particles hop to
a nearest neighbour empty site on the right with rate $q$. In addition
particles can hop from site $i$ on the upper lane to an empty site $i$
on the lower lane with rate $a$ and vice versa with rate $d$. To
analyze the phase diagram we note that the currents along the two
lanes are given by:
\begin{equation}
  J_\tau=p\tau (1-\tau)\quad;\quad J_\sigma= q \sigma (1-\sigma).
\end{equation}
Because of the exclusion on both lanes, the transverse current $K$ reads
\begin{equation}
  \label{eqn:KTASEP}
  K=a\sigma(1-\tau)-d\tau(1-\sigma).
\end{equation}
Using the $K=0$ condition, one can write the total current in an
equilibrated plateaux as a function of $\tau_0$ as 
\begin{equation}
  J_{\rm tot}=\tau_0(1-\tau_0) \left(p+q \frac {d a}{[a(1-\tau_0)+ d \tau_0]^2}\right).
\end{equation}
The total current as a function of $\tau_0$ is shown in
figure~\ref{fig:2TASEPFSCP} for {the most interesting case in
which} the curve has three extrema (depending on parameters there is also a less interesting case with only one extrema). This is the scenario we study below.

As before the stability of the plateaux changes when ${\rm d} J_{\rm
  tot} (\tau)/{\rm d} \tau =0$. As can be inferred from figure
\ref{fig:2TASEPFSCP} there are now three solutions which we denote,
from smallest to largest, by $\tau_M^1,\tau_m$ and $\tau_M^2$. These
lead to four distinct regimes of stability. When $\tau < \tau_M^1$ and
$\tau_m < \tau < \tau_M^2$ the plateau are diverging and when
$\tau_M^1 < \tau < \tau_m$ and $\tau > \tau_M^2$ the plateau are
converging. Using a procedure identical to the one outlined above one
can readily construct the phase diagram shown in
figure~\ref{fig:2TASEPFSPD}. {For example, consider phase $L$ in the
small $\tau_\ell$ region. One then notes that for $\tau_\ell <
\tau_M^1$ and $\tau_r < \tau_M^1$ there is a plateau controlled by the
left boundary with a small boundary layer on the right of the
system. When $\tau_\ell < \tau_M^1$ and $\tau_M^1 < \tau_r < \tau_m$
an upward shock connects the two boundaries. The system is then
controlled by the smaller current which is given by $J(\tau_l)$ up to
the line defined by $J(\tau_\ell)=J(\tau_r)$ and give rise again to a
density controlled by the left boundary. Similarly when $\tau_\ell <
\tau_M^1$ and $\tau_m < \tau_r < \tau_M^2$ the scenario is identical
but with the right plateau at $\tau_m$ and a small boundary layer at
the right end of the system. Again it is easy to see that the system's
density is controlled by $\tau_\ell$ up to the line defined by
$J(\tau_\ell)=J(\tau_m)$. Finally, when $\tau_\ell < \tau_M^1$ and
$\tau_r > \tau_M^2$ the stability analysis allows for two upward
shocks. One between $\tau_\ell$ and $\tau_m$ and one between $\tau_m $
and $\tau_r$. It is straightforward to see that the plateau with
$\tau_\ell$ dominates up to the line defined by
$J(\tau_r)=J(\tau_\ell)$.}

The same arguments can easily be extended to construct the rest of the
phase diagram shown in figure~\ref{fig:2TASEPFSPD}. The $L$ ($R$)
phase has a density controlled by the left (right) boundary and $mC$
is a minimal current phase with density $\tau_m$. The $MC_1$ ($MC_2$)
phase is a maximal current phase with a density $\tau_M^1$
($\tau_M^2$). Finally, noting that $J(\tau_M^1)=J(\tau_M^2)$ phase
$\overline{m}$ corresponds to a {\em shock phase} with a density $\tau_M^2$
on the left of the shock and a density $\tau_M^1$ on its right. Note
that there are two distinct phases controlled by the right boundary,
one is a high-density phase in which $\tau_r$ is larger than $\tau_M^2$
whereas the other is intermediate between $\tau_M^1$ and
$\tau_m$. Similarly, there are two distinct phases controlled by the
left boundary, one with a low density $\tau_\ell < \tau_M^1$ and one
with densities intermediate between $\tau_m$ and $\tau_M^2$.

\begin{figure}[ht]
  \begin{center}
    \includegraphics{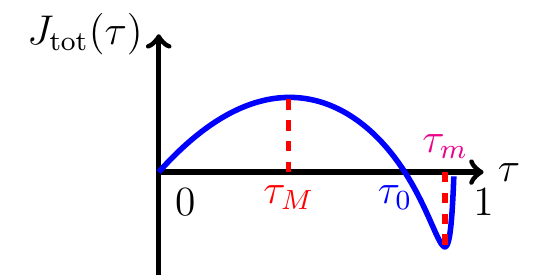} \includegraphics{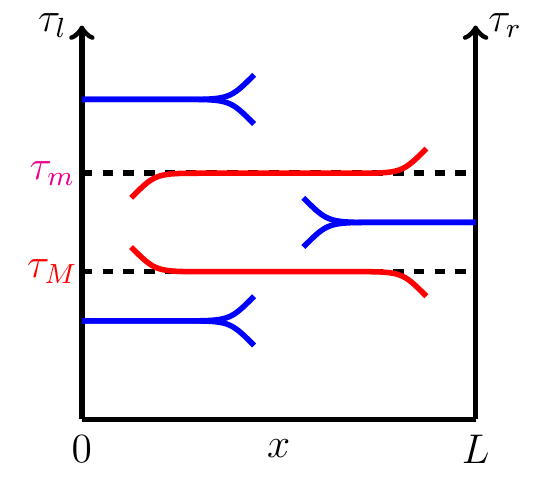}
  \end{center}
  \caption{Current density relation, typical stability profiles for 2 TASEPs flowing in opposite directions for $p=q=1$, $d=0.04$ and $a=1$. Note that the extremal densities have been spread out for clarity and their values are just illustrative.}
  \label{fig:2TASEPOFCP}
\end{figure}
  
\subsection{Two coupled TASEPs with advection in opposite directions}
It is natural to look at the phase diagram of two TASEP lattices when
the current in each of the lattices is directed in an opposite
direction. Specifically we consider the two TASEP model considered
above but where now particles on the upper lane hop to nearest
neighbour empty sites to their {\em left} with rate $q$.

Similarly to the case described above the currents on both lanes are
given by
\begin{equation}
  J_\tau=p\tau (1-\tau);\qquad J_\sigma= -q \sigma (1-\sigma),
\end{equation}
whereas the transverse current is still given by \eqref{eqn:KTASEP}. The $K=0$ condition now yields for the total
current $J_{\rm tot}$:
\begin{equation}
  J_{\rm tot}=\tau_0(1-\tau_0) \left(p- q\frac { d a }{[a (1-\tau_0)+ d \tau_0]^2}\right).
\end{equation}
In what follows we assume $a>d$ (the regime $a<d$ can be obtained by
symmetry) and furthermore, to obtain a phase diagram which is
significantly different from that of the single lane TASEP, we take
$pa/d > q > pd/a$. It is easy to check that when $q$ is outside this
range the current has only one extremum. The current in the regime of
interest is shown in figure~\ref{fig:2TASEPOFCP}. As is evident there
are now two points, denoted by $\tau_M$ and $\tau_m$, where the
stability changes. Plateaux are unstable for $\tau < \tau_M$ and $\tau
> \tau_m$ and are stable for $\tau_M < \tau < \tau_m$.

\begin{figure}[ht]
  \begin{center}
      \includegraphics{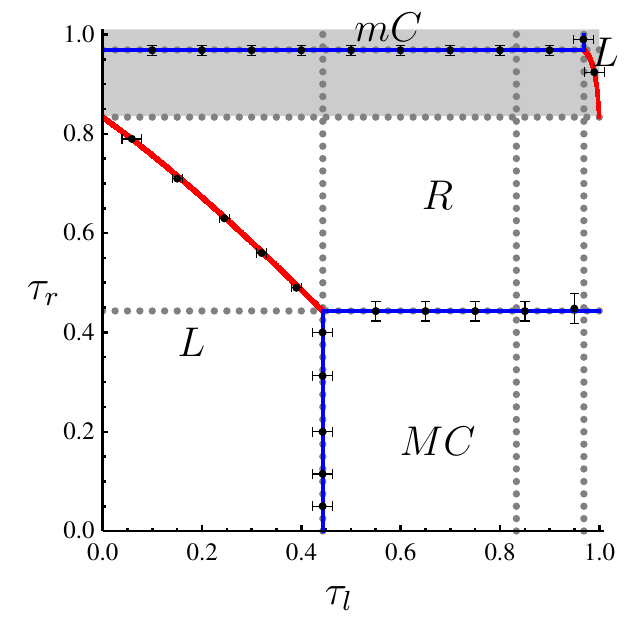}
  \end{center}
\caption{Phase diagram of 2 TASEPs with opposite advection
  directions. $p=q=1$, $d=0.04$ and $a=1$. The area with a negative
  current total current is greyed. {The dotted grey lines
    correspond to the extrema of the current while the black dots were
    obtained from continuous time Monte-Carlo simulations.}}
\label{fig:2TASEPOFPD}
\end{figure}

To construct the phase diagram we use the stability regimes (sketched
in figure~\ref{fig:2TASEPOFCP}) as discussed above. The results are
shown in figure~\ref{fig:2TASEPOFPD}. Phases $R$ and $L$ correspond to
phases which are controlled by the density of the right and left
reservoirs respectively. $MC$ corresponds to a maximal current phase
with a density $\tau_M$ and the phase $mC$ corresponds to a minimal
current phase with a density $\tau_m$. It is interesting to note that
inside phase $R$, say, as $\tau_r$ is increased, the {\em total}
current in the system changes direction. This, however, happens in a
{smooth} manner and is therefore {\it not} associated with a phase
transition in the system. Note that as in the previous cases, there
are two phases controlled by the left reservoirs, {corresponding} to low
and high density phases ($\tau_\ell < \tau_M$ and $\tau_\ell>\tau_m$,
respectively).

\section{A counter example---run-and-tumble model}
\label{app:WI}
The derivations presented throughout the paper rely on assumption
(ii) that the hopping rates of the particles within a lane depend only
on the occupancies of their site and neighbouring sites within their
lane.  In this section we show that when the hopping rates on each
lane depend on the occupancies of the others, implying interactions
between lanes and a violation of assumption (ii), the situation is
more subtle. In particular the results of~\ref{app:dynstab},
pertaining to the dynamical stability, do not always hold. To see
this, below we explicitly look at a specific two-lane model with
interactions.  We show that in this model the equilibrated plateaux
are not dynamically stable and therefore neither our stability
analysis nor the extremal current principle hold.

\begin{figure}[ht]
  \begin{center}
    \raisebox{1cm}{\includegraphics{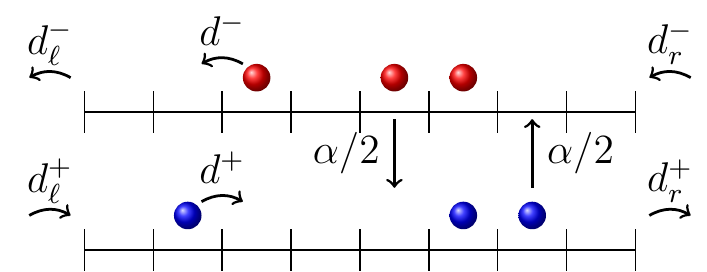}}\hspace{.5cm}\includegraphics{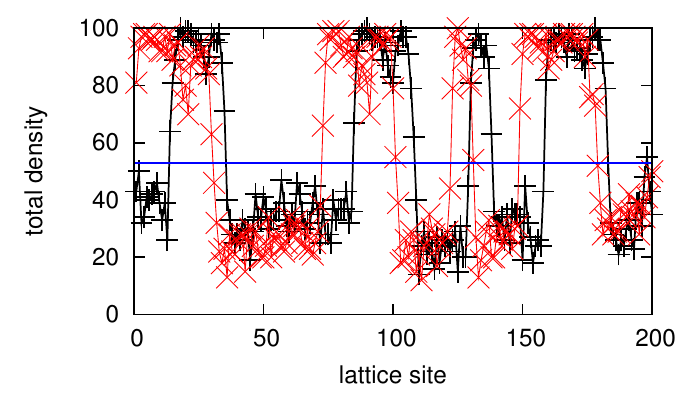}
  \end{center}
  \caption{{\bf Left:} Model of run-and-tumble particles on
    lattice. The interactions are set by the
    rates~\eqref{eqn:RTrates}. {\bf Right:} Result of continuous time
    Monte Carlo simulations for $\alpha=1$, $v^+=12$, $v^-=8$,
    $\rho=100$ and a system of $L=200$ sites. Both left and right
    reservoirs tend to impose plateaux of density
    $\tau_0=\sigma_0=26.5$. At $t=0$ the system is prepared with
    plateaux $\tau_0=\sigma_0=26$ but these plateaux are unstable and
    the system phase separates. Two successive snapshots of the total
    density $\tau_i+\sigma_i$ are plotted at $t=51$ (black) and $t=71$
    (red). This shows the advection of the domain walls.}
  \label{fig:modelapp}
\end{figure}

We consider a `run and tumble' model with partial exclusion (see
figure~\ref{fig:modelapp}).  Particles at site $i$ of the lower lane
hop to the right with rate $d_i^+$ while those on the upper lane hop
to the left with rate $d_i^-$. To account for partial exclusion the
rates $d^\pm_i$ at which particles hop are given by
\begin{equation}
  \label{eqn:RTrates}
  d_i^\pm = v^\pm \left(1-\frac{\tau_{i\pm 1}+\sigma_{i\pm 1}}{\rho}\right)
\end{equation}
These rates decrease with the occupancy of the arrival site and are
such that for any site $i$ the total occupancy $\tau_i+\sigma_i$ is
always lower than $\rho$ (which is treated as an external
parameter). In addition, particles switch lane (or `tumble') at rate
$\alpha/2$.

The symmetric case without reservoirs was studied in~\cite{TTCB2011}
where it was shown than flat profiles are unstable at high enough
density. The system undergoes phase-separation which results in an
alternating profile with low and high density plateaux, separated by
domain walls that perform random walks. The presence of advection
($v^+\neq v^-$) changes this picture slightly. At high enough density,
equilibrated plateaux are again unstable and tends to phase
separate. But as can be seen in figure~\ref{fig:modelapp}, the domain
walls {propagate with non-zero velocity, according to a
mass-conservation principle~\eqref{eqn:vshock}}.

Equilibrating the plateaux imposes $\tau_0=\sigma_0$ {and the
current-density relation is thus very similar to that of the TASEP}
\begin{equation}
  J(\tau_0)=(v^+-v^-) \tau_0\left(1-\frac{2 \tau_0}{\rho}\right)
\end{equation}
The phase shown in figure~\ref{fig:modelapp} has however no
counterpart in the TASEP phase diagram, which shows that the extremal
current principle does not apply here. In addition, the current
carried by the system can be shown to be unequal to the one imposed by
the boundaries.

\section{Conclusion}
\label{sec:conc}
We have introduced a stability analysis which can be used to derive
phase diagrams of driven diffusive systems.  In particular, we have
shown how the method can be applied to a class of two lane models and
demonstrated its applicability using several examples. In some of them
this allowed us to recover previously derived results with ease; we
have also mapped out new phase diagrams for more elaborate
models. Furthermore, the method was shown to be equivalent to a
extremal current principle and either method can be used according to
convenience.

{
The phase diagrams we have derived  reveal some interesting physics.
In the case of  a totally asymmetric exclusion process (TASEP)
coupled to a diffusive lane 
the phase diagram is unchanged from the single lane TASEP.
However when a bias is introduced into
the diffusive lane new phases emerge
such as a minimal current phase  
and both low and high density phases controlled by
the left boundary (figure 4). 
Similarly, when a TASEP
is coupled to another TASEP 
a rich phase diagram
including two left- and two right-boundary controlled phases,
two maximal current phases and a minimal current phase emerges
(figure 6).
In addition for two coupled TASEPs with currents in opposite directions,
a change in the direction of the total current may occur, although
there is no associated phase  transition (figure 8).
}

It is important to note that we have also pointed out that the
stability analysis, as well as the extremal current method, can fail
for a broader class of models than {outlined} in this paper. For
example, it can fail when the motion of particles on one lane is
influenced by the motion on the other lane. {To apply the method in
such cases the dynamical stability of plateaux has to be considered in
detail.}

Finally, we note that the method can be generalized to classes of
multi-lane models. This will be discussed in detail in a future
publication.

\appendix

\section{Dynamical Stability of Equilibrated Plateaux}
\label{app:dynstab}
In this appendix we show that for two-lane systems equilibrated
plateaux are always dynamically stable provided the hopping rate along
one lane does not depend on the occupancy of the other lane (a case
`without interactions' between the lanes). As shown in
section~\ref{app:WI} however, plateaux can be unstable if the
aforementioned condition is not satisfied.

Let us consider a perturbation around an equilibrated set of plateaux
$(\tau_0,\sigma_0)$:
\begin{equation}
  \tau(x)=\tau_0 + \sum_q \delta \tau_q {\rm e}^{i q x};\qquad  \sigma(x)=\sigma_0 + \sum_q \delta \sigma_q {\rm e}^{i q x}\;.
\end{equation}
At linear order in the perturbations, the Fourier modes decouple and
the mean-field equations~\eqref{eq:MFeqcont} yield
\begin{equation}
  \label{eqn:appLS1}
  \frac{\rm d}{{\rm d} t} \smallmatrix{\delta \tau_q\cr \delta \sigma_q\cr}= 
M \cdot \smallmatrix{\delta \tau_q \cr \delta \sigma_q};\qquad M=\smallmatrix{-D_\tau q^2 - i q J_\tau'+\partial_\tau K&\partial_\sigma K\cr
-\partial_\tau K&-D_\sigma q^2 -i q J_\sigma '-\partial_\sigma K\cr}\;,
\end{equation}
where $J_\tau'=\partial_\tau J_\tau$ and $J_\sigma'=\partial_\sigma
J_\sigma$. The plateaux are {\em dynamically}
stable if the eigenvalues of the matrix $M$ have
negative real parts. To see this, for simplicity, we first introduce $a,b,c,d$
such that
\begin{equation}
  M=\smallmatrix{a&b\cr
    c&d\cr}\;.
\end{equation}
The eigenvalues $\lambda^{\pm}$ are then given by
\begin{equation}
  2 \lambda^\pm = (a+d) \pm \sqrt{(a-d)^2+4 bc}\;.
\end{equation}
We denote by $R(z)$ and $I(z)$ the real and imaginary part of a complex
number $z$, respectively. Note that
\begin{equation}
  R(a+d)=-(D_\tau+D_\sigma) q^2-\partial_\sigma K+\partial_\tau K<0\;,
\end{equation}
since all the terms are negative. For the real part of $\lambda^\pm$
to be negative, it is thus sufficient to have
\begin{equation}
  \label{eqn:C1}
  |R(a+d)|>|R(\sqrt z)|\quad\text{with}\quad z=(a-d)^2+4 bc\;.
\end{equation}
Using the relation 
\begin{equation}
  R(\sqrt{z})=\sqrt{\frac{R(z)+|z|}2}\;
\end{equation}
we see that \eqref{eqn:C1} is equivalent to
\begin{equation}
  2 [R(a+d)]^2>R[(a-d)^2]+4bc+|z|\;,
\end{equation}
since $bc$ is real. Using this, a lengthy but straightforward algebra yields
\begin{equation}
  |a-d|^2-4 bc +8R (a) R(d)> |(a-d)^2+4 bc|\;.
\end{equation}
Furthermore, since
\begin{equation}
  R(a)R(d)=D_\tau D_\sigma q^4- \partial_\tau K \partial_\sigma K+q^2(\partial_\sigma K D_\tau - \partial_\tau K D_\sigma)>-\partial_\tau K \partial_\sigma K,
\end{equation}
we have $8 R(a) R(d)>8bc$ and thus
\begin{equation}
  |a-d|^2-4 bc +8 R(a) R(d)>  |a-d|^2+4 bc> |(a-d)^2+4 bc|\;,
\end{equation}
where the last inequality is the triangular inequality. Hence, the
real parts of the two eigenvalues are negative and the plateau is
{\em dynamically} stable.

\section{Eigenvalues and Eigenvectors of the Stationary Stability Problem}
\label{sec:appendixa}

In this appendix we present in more detail the stability of the
perturbations around equilibrated plateaux and the role they play for
the profiles. We first discuss in subsection \ref{app:A1} the
solution of equation \eqref{eqn:poleqx}
\begin{eqnarray}
  0&=  \lambda\, \chi(\lambda)\label{eqn:approots}\\
  \chi(\lambda)&= D_\tau D_\sigma \lambda^3 -\lambda^2(D_\tau \partial_\sigma J_\sigma+D_\sigma
  \partial_\tau J_\tau)  \label{eqn:appbouh}\\ 
  &\quad+\lambda(\partial_\sigma J_\sigma \partial_\tau
  J_\tau +D_\sigma
  \partial_\tau K-D_\tau \partial_\sigma K)\nonumber\\
  &\quad +\partial_\sigma
  K\partial_\tau(J_\tau)-\partial_\tau K
  \partial_\sigma(J_\sigma)\nonumber,
\end{eqnarray}
that controls the increase or decrease of the perturbations
{$(\epsilon,\eta)={\rm e}^{\lambda x} (\epsilon_0,\eta_0)$ that satisfy}
 \begin{equation}
   \label{eqn:appLS}
  0 = \smallmatrix{
    \lambda^2 D_\tau - \lambda \partial_\tau J_\tau +\partial_\tau K & \partial_\sigma K\cr
    -\partial_\tau K &      \lambda^2 D_\sigma-\lambda \partial_\sigma J_\sigma - \partial_\sigma K\cr} \cdot \smallmatrix{\epsilon_0\cr\eta_0\cr}.
\end{equation}
We then discuss in subsection \ref{app:A2} the stability analysis near
non-equilibrated reservoirs.
\subsection{Eigenvalues}
\label{app:A1}

First, one of the roots of \eqref{eqn:approots} is a trivial
$\lambda=0$ solution. The corresponding perturbation satisfies
\begin{equation}
(\epsilon_0,\eta_0) \propto
  (-\partial_\sigma K, \partial_\tau K).
\end{equation}
Such perturbation corresponds to shifting the values of the two
plateaux while keeping $K=0$. Indeed, one can readily check that
\begin{equation}
  {\rm d} K(\tau,\sigma)= \partial_\tau K \epsilon_0 + \partial_\sigma K
  \eta_0 \propto -\partial_\tau K \partial_\sigma K  + \partial_\sigma K
  \partial_\tau K=0.
\end{equation}

\begin{figure}[ht]
  \begin{center}
    \includegraphics{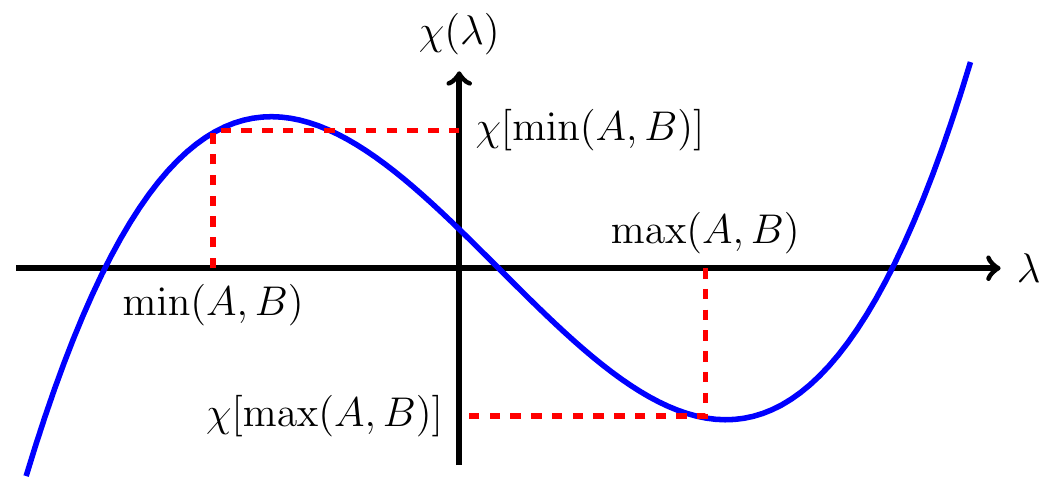}
  \end{center}
  \caption{Schematic plot of $\chi(\lambda)$. There is at most one
    root between $A$ and $B$.}
  \label{fig:Pol}
\end{figure}

We now show that there is only one solution $(\epsilon_0,\eta_0)$ of
equation \eqref{eqn:appLS} that satisfies $\epsilon_0  \eta_0 >0$. To see
this, we sum the two rows of equation \eqref{eqn:appLS} and divide by
$\lambda$ to get
\begin{equation}
  \lambda = \frac{D_\tau \epsilon_0}{D_\tau \epsilon_0+D_\sigma \eta_0} A + \frac{D_\sigma \eta_0}{D_\tau \epsilon_0+D_\sigma \eta_0}B,
\end{equation}
where
\begin{equation}
  A=\frac{\partial_\tau J_\tau}{D_\tau} \qquad\mbox{and}\qquad B=\frac{\partial_\sigma J_\sigma}{D_\sigma}.
\end{equation}
If $\epsilon_0$ and $\eta_0$ have the same sign, $\lambda$ has to lie
between $A$ and $B$. One however notes that 
\begin{equation}
  \chi(A)=(A-B) D_\sigma {\partial_\tau K};\qquad    \chi(B)=(A-B) D_\tau {\partial_\sigma K}.
\end{equation}
Since $\partial_\tau K<0$ and $\partial_\sigma K>0$, one then has that
\begin{itemize}
\item if $A>B$, $\chi(A)<0$ and $\chi(B)>0$
\item if $A<B$, $\chi(A)>0$ and $\chi(B)<0$,
\end{itemize}
and thus $\chi({\rm min}(A,B))>0$ and $\chi({\rm max}(A,B))<0$.  Since
the coefficient of $\lambda^3$ in $\chi(\lambda)$ is positive, there
is only one root lying between $A$ and $B$ (see figure
\ref{fig:Pol}). Since all this requires $\eta_0$ and $\epsilon_0$ to be of
the same sign, this means that there is only one root corresponding to
such a perturbation, thus completing our proof.

Note that for a marginal perturbation ($\lambda=0$) to exist one
needs $\chi(0)=0$, i.e.
\begin{equation}
  \label{appeqn:stabeq}
  \partial_\sigma K(\tau_0,\sigma_0)\partial_\tau J_\tau(\tau_0)-\partial_\tau K(\tau_0,\sigma_0) \partial_\sigma J_\sigma(\sigma_0)=0.
\end{equation}
This requires that $A$ and $B$ have opposite signs and thus the root
that has vanished lies between $A$ and $B$. This means that the only
root that can change sign is associated with
$\epsilon_0\,\eta_0>0$\footnote{Note that we have shown that the root
  associated to $\eta_0 \, \epsilon_0>0$ is the only one that can
  vanish. However, one has to be careful because the stability matrix
  is only diagonalisable at $\lambda \neq 0$ so that the study of the
  extremal current phases has to be done separately. In this case, the
  degenerate $\lambda=0$ eigenvalue is associated to a Jordan
  block. This is beyond the scope of this paper.}. 

We now consider the two other roots corresponding to $\epsilon_0\eta_0
<0$. They obey (see figure \ref{fig:Pol}) $\lambda_2< {\rm min}(A,B)$
and $\lambda_3> {\rm max}(A,B)$.  If $AB <0$ then clearly $\lambda_2<
0$ and $\lambda_3> 0$.  In the case {$AB >0$} we consider
\begin{equation}
\chi(0) = D_\tau A \partial_\sigma K - D_\sigma B \partial_\tau K\;.
\label{chi0}
\end{equation}
If $A>0$ and $B >0$, then $\lambda_1 >0$ and $\lambda_3 >0$
and, from (\ref{chi0}),
$\chi(0) >0$. Since $\chi(0)$ equals minus the product of the three roots
of the cubic equation $\chi(\lambda)=0$,
we deduce that $\lambda _2 <0$.
Similarly,
if $A<0$ and $B <0$, we have $\lambda_1 <0$ and $\lambda_2 <0$.
Then
$\chi(0) <0$ implies 
 $\lambda _3 >0$.
Thus, in all cases
 $\lambda _2 <0$ and $\lambda _3 >0$
and  the sign of $\chi(0)$  equals
that of $\lambda_1$.

\subsection{Connecting equilibrated plateaux to non-equilibrated reservoirs}
\label{app:A2}
In this section we show that the perturbations that satisfy $\eta_0
\epsilon_0<0$ play a crucial role in the boundary layers that connect
equilibrated plateaux to non-equilibrated reservoirs.

\begin{figure}[h]
  \begin{center}
    \includegraphics{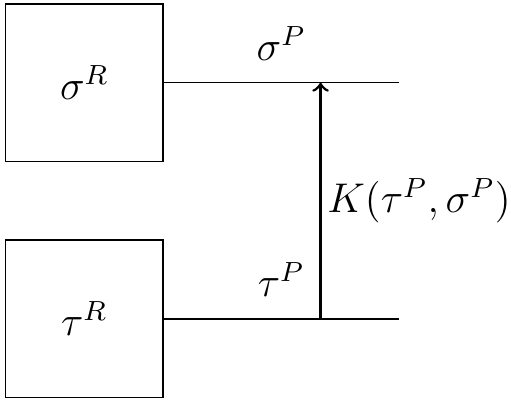}
    \caption{Schematic representation of the system}
    \label{fig:appres}
  \end{center}
\end{figure}

Let us consider the situation presented in figure \ref{fig:appres}.
The solution of the non-linear mean-field equations would lead to
relations between plateau and  reservoir densities
\begin{equation}
  \label{eqn:apH}
  \tau^P=H_1(\tau^R,\sigma^R)\quad;\quad\sigma^P=H_2(\tau^R,\sigma^R) \;,
\end{equation}
where the $R (P)$ superscript denotes densities at the reservoir (plateau). 
Since the plateaux densities are equilibrated, $H_1$ and $H_2$ satisfy
$K(H_1,H_2)=K(\rho_1^P,\rho_2^P)=0$. Considering arbitrary
perturbations $\delta \tau^R$ and $\delta\sigma^R$, the equilibrium
condition $(K=0)$ imposes
\begin{equation}
0=  \delta \rho_1^R [\partial_{\rho_1^P} K  \partial_{\rho_1^R} H_1 +\partial_{\rho_2^P} K 
  \partial_{\rho_1^R} H_2] + \delta \rho_2^R [ \partial_{\rho_1^P} K \partial_{\rho_2^R} H_1 + \partial_{\rho_2^P} K \partial_{\rho_2^R} H_2 ]
\end{equation}
Since this has to hold for any perturbation, one gets
\begin{eqnarray}
  \partial_{\rho_1^P} K  \partial_{\rho_1^R} H_1 +\partial_{\rho_2^P} K 
  \partial_{\rho_1^R} H_2&=&0\\
\partial_{\rho_1^P} K \partial_{\rho_2^R} H_1 + \partial_{\rho_2^P} K \partial_{\rho_2^R} H_2&=&0 
\end{eqnarray}
Last, this linear system of two equations for $\partial_{\rho_1^P} K$
and $\partial_{\rho_2^P} K$ has a non-trivial solution only if its
determinant vanishes, which requires:

\begin{equation}
  \label{eqn:relationH12}
{ \partial_{\rho_1^R} H_1 \partial_{\rho_2^R} H_2 - \partial_{\rho_1^R} H_2 \partial_{\rho_2^R} H_1 =0} \;.
\end{equation}

Next we construct a perturbation $\delta \tau^R, \delta \sigma^R$
which is constrained to leave the plateau densities unchanged, namely
$H_{1,2}(\tau^R+\delta\tau^R,\sigma^R+\delta\sigma^R)=H_{1,2}(\tau^R,\sigma^R)$.
This requires
\begin{eqnarray}
  \label{eqn:apptoto}0&=&\partial_{\tau^R} H_1 \delta \tau^R +\partial_{\sigma^R} H_1
  \delta \sigma^R \\   \label{eqn:apptata} 0&=&\partial_{\tau^R} H_2 \delta \tau^R
  +\partial_{\sigma^R} H_2 \delta \sigma^R 
\end{eqnarray}
Again, this linear system admits a non-trivial solution iif the
determinant vanishes so that
\begin{equation}
  \partial_{\tau^R} H_1 \partial_{\sigma^R}H_2-\partial_{\tau^R}H_2 \partial_{\sigma^R}H_1 =0 \;,
\end{equation}
which holds due to Eq.~\eqref{eqn:relationH12}. Finally,
equations \eqref{eqn:apptoto} and \eqref{eqn:apptata} show that the
perturbation one has to look for is of the form
\begin{equation}
  \label{eqn:resper1}
  (\delta \tau^R,\delta \sigma^R) \propto (-\partial_{\sigma^R} H_1, \partial_{\tau^R} H_1)
\end{equation}
or equivalently
\begin{equation}
  \label{eqn:resper2}
  (\delta \tau^R,\delta \sigma^R) \propto (\partial_{\sigma^R} H_2, -\partial_{\tau^R} H_2)
\end{equation}
This in particular means that $\delta \tau^R \delta \sigma^R
<0$. Close to the left boundaries, such perturbations have to decay
when $x$ increases and they correspond to the eigenvector with
components of opposite signs that has $\lambda<0$ as discussed in the
previous section. Conversely, the same reasoning for the right
boundaries leads relevant perturbations with $\delta \tau^R
\delta \sigma^R <0$ and $\lambda>0$. This corresponds to the eigenvector
with components of opposite sign and $\lambda>0$ as discussed in the
previous section.

Concluding, for general systems with non-equilibrated reservoirs,
there are boundary layers connecting the reservoirs to equilibrated
bulk plateaux, whose behavior is dictated by the eigenvectors with
$\epsilon_0 \eta_0<0$, one for each boundary. The connection between the
two plateaux is then dictated by the sign of the remaining eigenvalue,
that corresponds to $\epsilon_0 \eta_0>0$.

\section{Connection with the Extremal Current Principle}
\label{sec:appMSA}
For single lane driven diffusive systems, studies based on domain wall
theory have led to the formulation of the `extremal current
principle'~\cite{PS99}. In the multilane cases, it is
also possible to show that for the transverse current $K$ satisfying
condition~\eqref{eqn:Keq} and in the absence of interactions between
the lanes\footnote{namely the hopping rate in one lane is independent
  of the occupancy on the other lane}, our stability analysis shows
that an `extremal current principle' applies.

Namely, for {\em increasing} profiles ($\tau_\ell \leq \tau_r$) the
steady profiles form an equilibrated set of plateaux such that $J_{\rm
  tot}(\tau)$ realizes its {\em minimum} for $\tau \in
[\tau_\ell,\tau_r]$. Conversely, for {\em decreasing} profiles
($\tau_\ell > \tau_r$), the steady profiles form an equilibrated set
of plateaux such that $J_{\rm tot}(\tau)$ realizes its {\em maximum}
for $\tau \in [\tau_r,\tau_\ell]$. The demonstration of both assertions
being almost identical, we only consider here the {\em increasing}
case.

There are four cases to consider depending on whether $\tau_\ell$ and
$\tau_r$ correspond to stable or unstable plateaux, that is satisfy
$J_{\rm tot}'<0$ or $J_{\rm tot}'>0$ (here quotes abbreviate total
derivative with respect to $\tau$). Maximal current profiles are
irrelevant since they correspond to {\em decreasing} profiles but
minimal current profiles play an important role since they correspond
to {\em increasing} profiles. When they exist, we denote by $\tau_m^1$
... $\tau_m^k$ the densities for which $J_{\rm tot}$ has local maxima.

\paragraph{(i) $J_{\rm tot}'(\tau_\ell)>0$ and $J_{\rm tot}'(\tau_r)>0$.}

If $J_{\rm tot}'(\tau)$ remains strictly positive on the interval
$[\tau_\ell,\tau_r]$, $\tau_\ell$ and $\tau_r$ are not separated by a
minimal current phase and the steady-state profile corresponds to
$\tau(x)=\tau_\ell$ with a short boundary layer on the right. Since
$J_{\rm tot}'(\tau)>0$ on $[\tau_\ell,\tau_r]$, the steady-state
profile indeed minimize $J_{\rm tot}$ on the interval
$[\tau_\ell,\tau_r]$.

If $J_{\rm tot}'$ vanishes on $[\tau_\ell,\tau_r]$, the end-points are
separated by at least one minimal current phase. A profile consistent
with the stability analysis is given by a sequence of plateaux of
densities in $\{\tau_\ell,\tau_m^i\}$ connected by up-going
shocks. Mass conservation -- equation \eqref{eqn:vshock} -- shows that
an up-going shock separating two density plateaux propagates towards
the plateau with the largest current so that the plateaux with the
smallest value of the current spread while the others recede. Since
$J_{\rm tot}'(\tau_\ell)>0$ and $J_{\rm tot}'(\tau_r)>0$, the smallest
value among $\{J_{\rm tot}(\tau_\ell), J_{\rm tot}(\tau_m^i)\}$ is
indeed the global minimum of $J_{\rm tot}$ in $[\tau_\ell,\tau_r]$.

\paragraph{(ii) $J_{\rm tot}'(\tau_\ell)>0$ and $J_{\rm tot}'(\tau_r)<0$.}
$\tau_\ell$ is unstable and $\tau_r$ is stable. A profile consistent
with the stability analysis is given by a sequence of plateaux of
densities in $\{\tau_\ell,\, \tau_m^i,\,\tau_r\}$ connected by
up-going shocks. Again, the plateaux with the smallest current spread
while the others recede and the observed value of $J_{\rm tot}$
is its global minimum in $[\tau_\ell,\tau_r]$.

\paragraph{(iii) $J_{\rm tot}'(\tau_\ell)<0$ and $J_{\rm tot}'(\tau_r)<0$.}

$\tau_\ell$ and $\tau_r$ are stable. If there are no local minimum
in $[\tau_\ell,\tau_r]$, the steady profile is a plateau at
$\tau=\tau_r$, $J_{\rm tot}'$ remains strictly negative in
$[\tau_\ell,\tau_r]$ and the current is indeed minimized over
$[\tau_\ell,\tau_r]$.

If $J_{\rm tot}'$ changes sign, it has {to do so} at least twice and
there are local minima $\tau_m^i$ between $\tau_\ell$ and $\tau_r$. A sequence of
plateau of increasing densities $\tau(x)\in\{\tau_m^i,\tau_r\}$
connected by shocks is consistent with the stability analysis. Again,
the plateaux with the lowest current spread while the others {recede}
and the corresponding value of the current is the global minimum of
$J_{\rm tot}$ in $[\tau_\ell,\tau_r]$.

\paragraph{(iv) $J_{\rm tot}'(\tau_\ell)<0$ and $J_{\rm tot}'(\tau_r)>0$.}

There is at least one minimum $\tau_m^i$ between $\tau_\ell$ and
$\tau_r$.  A sequence of plateau of increasing densities
$\tau(x)=\tau_m^i$ is consistent with the stability analysis. If there
are several minima, the plateaux with the lowest value of the current
spread while the others {recede}. Since $J_{\rm tot}'(\tau_\ell)<0$ and
$J_{\rm tot}'(\tau_r)>0$, the global minimum of $J_{\rm tot}$ is realized by one
of the intermediate minimum and the value of $J_{\rm tot}$ is thus
indeed minimized on the whole interval.

This concludes the list of 4 possible cases and shows that the current
is always minimized between $\tau_\ell$ and $\tau_r$. Note that in all
the above discussion, the minima of the current can be realized by
several densities simultaneously. This leads to the coexistence of
several plateaux with different densities.

For {\em decreasing} profiles, the discussion is very similar, with
two main differences. First minimal current profiles do not play any
role while maximal current ones do, since only the latter are {\em
  decreasing}. Then, a downwards shock propagates towards the plateaux
with the {\em smallest} current, so that the plateaux with the
largest current spread.

\end{document}